\def\cm2{cm$^{-2}$}

\def\c2{C~{\sc ii}}
\def\c4{C~{\sc iv}}
\def\fe2{Fe~{\sc ii}}
\def\fe3{Fe~{\sc iii}}
\def\mg1{Mg~{\sc i}}
\def\mg2{Mg~{\sc ii}}
\def\si2{Si~{\sc ii}}
\def\si4{Si~{\sc iv}}
\def\al2{Al~{\sc ii}}
\def\al3{Al~{\sc iii}}
\def\o1{O~{\sc i}}
\def\n1{N~{\sc i}}
\def\h1{H~{\sc i}}

\def\teff{\mbox{T$_{\rm eff}$}}
\def\logg{\mbox{log~{\it g}}}
\def\vmicro{\mbox{$v_{\rm t}$}}
\def\vsini{\mbox{$V$sin$i$}}
\def\kmsec{\mbox{km~s$^{\rm -1}$}}
\def\approxlt{\mathrel{\spose{\lower 3pt\hbox{$\sim$}}
        \raise 2.0pt\hbox{$<$}}}
\def\approxgt{\mathrel{\spose{\lower 3pt\hbox{$\sim$}}
        \raise 2.0pt\hbox{$>$}}}

\documentclass[article]{has40}
\usepackage{graphicx}
\usepackage{amssymb}
\tabletypesize{\normalsize}  
\usepackage{float}
\def\plotone#1{\centering \leavevmode
\includegraphics[width=.95\columnwidth]{#1}}

\def\plotone#1{\centering \leavevmode
\includegraphics[width=.95\columnwidth]{#1}}

\shortauthors{Govea et al.}
\shorttitle{RRc star AS162158}

\begin{document}
\large    
\pagenumbering{arabic}
\setcounter{page}{188}

\title{Chemical Abundances of RR Lyrae Type C Star AS162158}

\author{
Jose Govea\altaffilmark{1},
Thomas Gomez\altaffilmark{1},
George W. Preston\altaffilmark{2},
Christopher Sneden\altaffilmark{1}
}

\altaffiltext{1}{Department of Astronomy and McDonald Observatory, 
                 The University of Texas, Austin, TX 78712; 
                 jgovea@utexas.edu, gomezt@astro.as.utexas.edu, chris@verdi.as.utexas.edu}

\altaffiltext{2}{The Observatories of the Carnegie Institution of 
                 Washington, Pasadena, CA 91101; iii@ociw.edu}

\begin{abstract}
We report the first extensive model atmosphere and detailed chemical 
abundance study of eight RR Lyrae variable stars of c subclass 
throughout their pulsation cycles. 
Atmospheric parameters effective temperature, surface gravity, 
microturbulent velocity, and metallicity have been derived. 
Spectra for this abundace analysis have been obtained with the echelle 
spectrograph of 100-inch du Pont telescope at Las Campanas Observatory. 
We have found metallicities and element abundance ratios to be constant
within observational uncertainties at all phases of all stars. 
Moreover, the $\alpha$-element and Fe-group abundance ratios with respect
to iron are consistent with other horizontal-branch members (RRab, blue
and red non-variables). The [Fe/H] values of these eight RRc  stars
have been used to anchor the metallicity scale of a much larger sample
of RRc stars obtained with low S/N ``snapshot'' spectra.

\end{abstract}

\section{Introduction}
The H-R Diagram horizontal branch is populated with low-metallicity stars 
that are doing quiescent core helium fusion.
The instability strip portion of the horizontal branch contains RR Lyrae 
stars, which are radially pulsating giant A-F stars. 
RRc stars occupy the hotter
portion of the instability strip while RRab occupy the cooler end.
Generally, RRc stars have effective temperatures in the range 7000K-7500K
while RRab stars are cooler, with $\teff$ ranges from 6000K-7000K. 

RR Lyrae stars have been used as standard candles to measure galactic
distances through photometry and low-resolution spectroscopy because
of their uniform brightness. 
They are also potentally excellent tracers of chemical properties of old 
stellar populations. 
But high-resolution spectroscopic analyses of RR~Lyr are not plentiful 
because their short pulsation periods ($\sim0.1-1.0$ days) and large 
radial-velocity ($RV$) amplitudes impose limits on exposure times, 
and consequent signal-to-noise ($S/N$) ratios in their spectra. 
The RRc variables have been neglected in these high-resolution 
studies; there are fewer RRc's than RRab's and most are somewhat
faint for high-resolution spectroscopy in the past. 

Seven of our stars exhibit the so-called Blazhko effect, a slow modulation
of their basic pulsational periods; Blazhko periods of these stars are
less than 12 days. 
This period is small enough to produce detectable
rotational broadening (\vsini~$>$ 15 \kmsec) of their spectral lines via 
$PV$~=~$2 \pi R$, in which $P$, $V$, and $R$ are the 
rotation period, axial rotational velocity, and radius of the star. 
In this paper we focus on one of our program stars, AS162158, showing
the spectroscopic changes that occur throughout it's pulsation cycle. 
Further details regarding the whole set of the program stars are given
in Govea et al. (2013)\nocite{gov13}.

\section{Model Atmosphere and Abundace Determinations}

What is of concern here are the line spectrum cyclic changes that 
AS162158 undegoes throughout its pulsational cycle. 
We imposed a 20-minute integration limit on the spectroscopic observations 
that, once normalized, yielded low $S/N$ values ($<$50).  
In order to prepare the spectra for equivalent width ($EW$) measurements
and subsequent chemical composition analysis, we combined the spectra 
in narrow phase intervals to increase mean $S/N$ ratios. 
We created narrow phase intervals that were grouped into phase bins 
no larger than $\Delta\phi$~$\simeq$~0.05 to minimize contamination 
(e.g., spectral line velocity smearing) due to the rapid atmospheric changes. 
Once the spectra were normalized we applied a Gaussian line profile 
fitting to measure equivalent widths. 

We acquired model atmosphere parameters \teff, \logg, \vmicro, 
[Fe/H]\footnote{
We adopt the standard spectroscopic notation (Helfer, Wallerstein,
\& Greenstein 1959\nocite{hel59}) that for elements A and B,
[A/B] $\equiv$ log$_{\rm 10}$(N$_{\rm A}$/N$_{\rm B}$)$_{\star}$ $-$
log$_{\rm 10}$(N$_{\rm A}$/N$_{\rm B}$)$_{\odot}$.}
and relative abundance ratios [X/Fe] for our program stars at
each of their co-added phases. 
Phase elemental abundances were acquired using the latest version of
local thermodynamic equilibrium spectral line synthesis code MOOG 
(\citealt{sne73})\footnote{
Available at http://www.as.utexas.edu/~chris/moog.html}. 
Model atmospheres were interpolated from the ATLAS grid 
(\citealt{kur92})\footnote{
Available at http://kurucz.harvard.edu/grids.html}
which were calculated  assuming $\alpha$-element enhancements
and opacity distribution functions (\citealt{cas03}). 
Trial Kurucz model atmospheres and $EW$ line lists were input 
parameters for MOOG, whose output was individual line abundances from
iterative force-fitting of predicted $EW$s to the measured values. 
We adjusted input model parameters in a similar fashion as \cite{for11a}. 
To summarize, $\teff$ was adjusted until there was no signicant trend between abundances 
and excitation energies of the {\ion{Fe}{1} lines. 
Values for $\logg$ needed to agree with observational errors of
the derived 
abundances from \ion{Fe}{1} and \ion{Fe}{2} lines. \ion{Fe}{1} and \ion{Fe}{2} lines conceded \vmicro values
when there was a lack of abundace trend. 
Finally, [M/H] metallicity parameter was obtained by derived Fe abundances.  

\begin{center}
\begin{deluxetable}{ccccccccccc} 
\tabletypesize{\footnotesize} 
\tablewidth{0pt} 
\tablecaption{Atmospheric Parameters for AS162158\label{as162158tab}}
\tablecolumns{11} 
\tablehead{
\colhead{$\phi$}         & 
\colhead{\teff}             &  
\colhead{log $g$}        &
\colhead{$v_t$}            &
\colhead{[M/H]}           &
\colhead{[Fe/H]}          &
\colhead{\#lines}         &
\colhead{$\sigma$}     &
\colhead{[Fe/H]}          & 
\colhead{\#lines}         &
\colhead{$\sigma$}     \\
\colhead{}                   & 
\colhead{K}                 &  
\colhead{}                   &
\colhead{\kmsec}       &
\colhead{}                   &
\colhead{\ion{Fe}{1}}   &
\colhead{}                   &
\colhead{}                   &
\colhead{\ion{Fe}{2}}   & 
\colhead{}                   & 
\colhead{}                
}
\startdata 

 0.039 &       7400 &       2.10 &       2.40 &    $-$2.00 &    $-$1.96 &         14 &       0.09 &    $-$1.91 &         18 &       0.17 \\

 0.138 &       7300 &       2.10 &      2.00 &    $-$2.00 &    $-$1.88 &         11 &       0.16 &    $-$1.86 &         14 &       0.22 \\ 

 0.225 &       7100 &       2.30 &       2.10 &    $-$2.00 &    $-$1.83 &         13 &       0.17 &    $-$1.80 &         14 &       0.19 \\

 0.405 &       7000 &       2.20 &       2.00 &    $-$2.00 &    $-$1.72 &         27 &       0.27 &    $-$1.76 &         15 &       0.30 \\

 0.631 &       7100 &       2.70 &       2.40 &    $-$2.00 &    $-$1.77 &         44 &       0.25 &    $-$1.83 &         22 &       0.13 \\

 0.863 &       7400 &       2.40 &       2.40 &    $-$2.00 &    $-$1.79 &         17 &       0.29 &    $-$1.83 &         17 &       0.19 \\
\enddata
\end{deluxetable}
\end{center}
 
Table~\ref{as162158tab} shows the variation of atmospheric models derived
for AS162158 along with [Fe/H] metallicities.
We find that atmospheric parameters change in expected ways. 
That is, the hottest temperatures and highest gravities occur near
phase $\phi \sim$ 0.5, at which time the photometric brightness of an
RRL is at minimum and derived radial velocity is at maximum.
However, in spite of the 400K $\teff$ variation and 0.4~dex \logg\ variation
there is only a small variation in derived [Fe/H], especially as
indicated by the Fe~II lines.

In Figure~\ref{fig1} we show the variations on the relative [X/Fe]
abundance ratios for six lighter elements in AS162158.
The dotted line in each panel represents the average abundance of the 
element throughout the pulsation cycle. 
Inspection of this figure suggests that derived abundance ratios for 
these elements (a) are essentially independent of pulsational phase,
and (b) are consistent with those of other members of the metal-poor
horizontal branch (RHB, RRab, and BHB stars). 
In particular, all of the $\alpha$ and $\alpha$-like elements (Mg, Si, Ca
and Ti) are enhanced, [X/Fe]~$\simeq$ $+$0.3 to $+$0.5, 
just as they are in other metal-poor stars.
This result holds generally for all of the RRc stars in the \cite{gov13}
sample.

\begin{figure}[H]
\epsscale{0.35}
\plotone{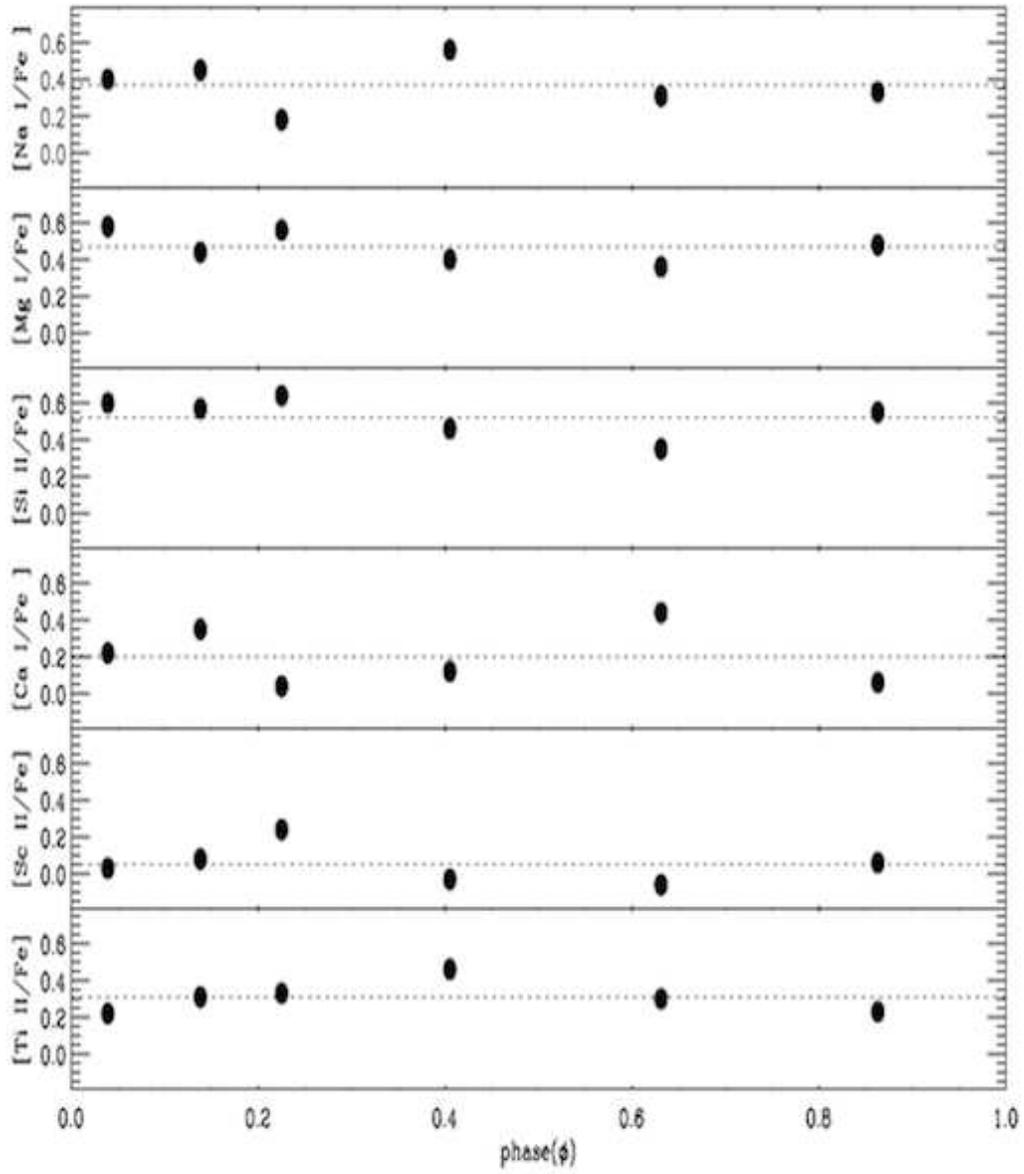}
\caption{\footnotesize Relative abundance ratios for lighter elements as 
   functions of phase for AS162158. 
   Dotted line represents the average chemical abundance. 
   \label{fig1}} 
\end{figure}

 \pagebreak
\section{Conclusions}
We have calculated model atmospheric parameters, metallicity, and 
abundance ratios for eight field RRc variable stars. 
In this paper we narrowed in on AS162158. 
Further details concerning the rest of the program stars are given in
\cite{gov13}. 
Each star in our survey has been explored at nearly all pulsational phases 
by taking short exposure times and co-adding spectra. 
Deriving \teff, \logg, \vmicro, and [Fe/H] for AS162158 based on 
spectroscopic constraints, we find that (a) the mean $\teff$  for AS162158 
is approximately 7200K, the mean apparent gravity is about $\logg=2.2$, 
while the obtained metallicity value is [Fe/H]=-2.00. 
Additionally, as detailed in the full paper, the mean microturbulent
velocities are \vmicro~$\simeq$~2.2~km/sec.
These velocities are much lower than those derived by \cite{for11b} for
the cooler RRab stars, and are consistent with the overall smaller photometric
and spectroscopic parameter variations of the RRc stars than their RRab 
counterparts.
We also find that the $\alpha$-element (Mg, Si, Ca, Ti) abundances 
on average are overabundant ([X/Fe]~$\simeq$~$+$0.3 to 0.4)
and the Fe-group (Sc, Cr, Ni) have solar abundance ratios ([X/Fe]~$\simeq$~0)
just as they are in other halo population samples.
Further implications of our study are detailed in \cite{gov13}, including
the use our program stars's [Fe/H] values to calibrate
the metallicity estimates of a large-sample RRc snapshot           
spectroscopic survey of \cite{kol13}.

\section{Acknowledgements}
Our investigations of RR Lyrae stars have been supported by 
NSF grants AST-0908978 and AST-1211585 and by the Baker Centennial 
Research Endowment to the Astronomy Department of the University of Texas.

\end{document}